\documentstyle[preprint,prc,aps]{revtex}

\begin{document}


\preprint{OSU--95-101}

\title{The Gluon Condensate and Running Coupling of QCD}

\author{Hua-Bin Tang and R. J. Furnstahl}
\address{Department of Physics\\
         The Ohio State University,\ \ Columbus,~Ohio\ \ 43210}
\date{February 16, 1995}

\maketitle

\begin{abstract}
An expression for the photon condensate in quantum electrodynamics
is presented and generalized to deduce  a simple relation between
the gluon condensate and the running coupling
constant of quantum chromodynamics (QCD). Ambiguities in defining the
condensates are discussed. The values of the
gluon condensate from some Ans\"{a}tze for the running
coupling in the literature are compared with the value
determined from QCD sum rules.

\vspace{10mm}

\end{abstract}

\tighten
\newpage

The running coupling constant of
quantum chromodynamics (QCD) in the low-momentum region is so far
scarcely known, and yet it is associated with various approaches to
low-energy nonperturbative QCD.
For example, in the Dyson-Schwinger equation (DSE)\cite{CRAIG93}
approach, one usually assumes some Ansatz for the running coupling
constant in momentum space to approximate the gluon propagator.
In quark potential models,
one also needs to know the associated interaction potential.
Thus, apart from the well-known constraints from confinement and
asymptotic freedom, additional
constraints on the running coupling constant in the low-momentum
region would be very useful. One such constraint is discussed here.

The phenomenological success of QCD sum rules\cite{SHIFMAN92} has
stimulated many investigations
\cite{QUINN82,DAVID82,TAYLOR83,NOVIKOV85,DAVID86,BROWN89}
on the validity of
the operator-product expansion (OPE)\cite{WILSON69}.
An important result has been the realization
that one should introduce a renormalization scale
$\mu$ in the Wilson coefficients as well as the
condensates\cite{NOVIKOV85,MUELLER85}.
The high-momentum $(p>\mu$ with $p$ being Euclidean)
parts of the operator product are
put into the  coefficients and
the low-momentum $(p<\mu)$ parts are parametrized by the condensates.
Although the formal validity of the OPE is now generally accepted,
there are still many discussions\cite{MUELLER93,GRUNBERG94,JI94}
concerning the practical
separation of the low- and high-momentum contributions and
the ambiguities associated with the
infrared renormalon singularities\cite{THOOFT77,PARISI78,MUELLER85}.

A related issue is whether one can separate the perturbative
contributions in the form of powers of $\mu$ from the so-called
``genuine'' nonperturbative effects in the condensates.
While Ref.~\cite{NOVIKOV85} suggests that the separation is possible,
Ref.~\cite{DAVID86} seems to imply the opposite.
In this Letter, we hope to clarify the situation by
using some simple examples.
We proceed to the discussion of the photon
condensate in quantum electrodynamics (QED). Generalizing this QED
result, we deduce
a simple relation between the
gluon condensate and the running coupling constant of QCD.
Some Ans\"{a}tze for the running coupling in the literature are
then examined.

We start with the simplest example
of the composite operator $\phi^2$ with $\phi(x)$ being a {\it free}
scalar field.
One can define $\phi^2$ as the $x\rightarrow 0$
limit of the regular part of $T[\phi(x)\phi(0)]$:
\begin{equation}
\phi^2   \equiv \lim_{x\rightarrow 0}
       \Bigl [T[\phi(x)\phi(0)]-S(x)\Bigr]      \ , \label{eq:limit}
\end{equation}
where $T$ stands for the usual time-ordering and $S(x)$
is the part of $T[\phi(x)\phi(0)]$ that is singular at $x=0$
and it can contain other composite operators with singular
coefficients when interactions are present.
In general $S(x)$ cannot be uniquely defined since, at the very least,
it may contain a regular function multiplied by the singular logarithm
$\ln(\mu^2x^2)$, where $\mu$ is some arbitrary scale to make the
argument dimensionless.  We call $S(x)$ {\it minimal}
when the regular parts of  $S(x)$ come {\it only} from their multiplication
with the singular function $\ln(\mu^2x^2)$ but not with powers of $\mu$.
It is instructive to compare the vacuum matrix element of $\phi^2$
in cutoff and dimensional regularizations with $S(x)$ being minimal.

Let's first consider
using cutoff regularization (CR). We can write
\begin{eqnarray}
\langle 0|T[\phi(x)\phi(0)]|0\rangle  & = &
     \int_{k > \mu} {{\rm d}^4 k \over (2\pi)^4}\,
      e^{ik\cdot x} \Bigl( {1\over k^2}-{M^2\over k^4}
         +{M^4\over k^6}+\cdots  \Bigr)
        \nonumber \\
  & &    + \int_{k < \mu}  {{\rm d}^4 k \over (2\pi)^4}\,
        {1 \over  k^2+M^2}
        \Bigl[ 1+ik\cdot x-{1\over 2}(k\cdot x)^2+\cdots \Bigr]
                                   \ . \label{eq:pgtCR}
\end{eqnarray}
where $M$ is the scalar mass and we have introduced
the cutoff scale $\mu$ with $\mu \gg M$. The momentum is Euclidean.
Notice that the first integral is not a
minimal singular part.
Indeed performing the integral term-by-term one sees that
it contains the regular piece
\begin{equation}
-{M^2\over 16\pi^2}\Bigl({\mu^2\over M^2}+1-2\gamma+\ln 4\Bigr)
        +{M^4\over 128\pi^2}
      \Bigl( {\mu^4 \over 2M^4}
        -{\mu^2 \over M^2}-4\gamma+3+2\ln 4\Bigr)x^2+\cdots \ ,
\end{equation}
where $\gamma=0.577215$ is Euler's constant.
The above smooth piece should be removed  from the first integral
to make it minimal.
After such a removal and using Eq.~(\ref{eq:limit}), we obtain
\begin{eqnarray}
\langle 0|\phi^2|0\rangle   &=&
   \int_{k<\mu}\! {{\rm d}^4 k \over (2\pi)^4}\,
        {1 \over  k^2+M^2} -
       {M^2\over 16\pi^2}\Bigl({\mu^2\over M^2}+1-2\gamma+\ln 4\Bigr)
                 \nonumber \\
       & = &     -{M^2 \over 16\pi^2}
            \Bigl(\ln {\mu^2\over M^2}+1-2\gamma+\ln 4\Bigr)
         \ . \label{eq:CR}
\end{eqnarray}

If we consider dimensional regularization (DR) instead,
we can also make an analogous separation,
\begin{eqnarray}
\langle 0|T[\phi(x)\phi(0)]|0\rangle  & = &
      \mu^{4-d}\sum_{n=0}^{\infty}\int {{\rm d}^d k \over (2\pi)^d}\,
      e^{ik\cdot x} {(-1)^n M^{2n}\over (k^2)^{n+1}}
       \nonumber \\
   & &    + \mu^{4-d}\sum_{n=0}^{\infty}
         {1\over n!}\int  {{\rm d}^d k \over (2\pi)^d}\,
        {(ik\cdot x)^n \over  k^2+M^2}
                                      \ , \label{eq:pgtDR}
\end{eqnarray}
where
the first sum contains the singular piece\cite{LIGHTHILL70}.
Eq~(\ref{eq:pgtDR}) is justified as follows. Subtracting
the first (singular) sum on the right-hand side
of Eq~(\ref{eq:pgtDR}) from the propagator on the left-hand side,
expanding  the exponential $e^{ik\cdot x}$ in the resulting regular
expression, and using
\begin{equation}
    \int {\rm d}^d k (k^2)^\beta = 0  \label{Dint}
\end{equation}
with arbitrary $\beta$, one arrives at the second sum.%
\footnote{It may be a helpful
exercise for skeptical readers to explicitly check
Eq~(\ref{eq:pgtDR}) for d=1 when all the integrals
are trivial. In that case, the separation in Eq~(\ref{eq:pgtDR})
is into a {\it purely} singular piece:
$-(|x|/2)\sum_{n=0}^{\infty}(M^2x^2)^n/(2n+1)!$
and a regular piece:
$(1/2M)\sum_{n=0}^{\infty}(M^2x^2)^n/(2n)!$.}
The terms in the second sum of Eq~(\ref{eq:pgtDR})
are the so-called non-singular parts
in the OPE and are proportional to $\delta$-functions and their
derivatives in momentum space\cite{SHIFMAN92}. Notice
again that the first sum in Eq~(\ref{eq:pgtDR}) is not minimal.
Performing the integrals and letting
$d \rightarrow 4$, one finds that
the following regular Taylor series (singular in $d$)
\begin{equation}
-{M^2\over 16\pi^2}\Bigl({2\over d-4}-\gamma-\ln \pi\Bigr)
        -{M^4\over 128\pi^2}
      \Bigl({2\over d-4}-\gamma+1-\ln \pi\Bigr)x^2+\cdots
\end{equation}
should be removed to make the singular integral minimal.
Eq.~(\ref{eq:limit}) then leads to
\begin{eqnarray}
\langle 0|\phi^2|0\rangle   &=& \mu^{4-d}\!
   \int\! {{\rm d}^{d} k \over (2\pi)^{d}}\,
        {1 \over  k^2+M^2} -
       {M^2\over 16\pi^2}\Bigl({2\over d-4}-\gamma-\ln \pi\Bigr)
        \nonumber \\
       & = &     -{M^2 \over 16\pi^2}
            \Bigl(\ln {\mu^2\over M^2}+1-2\gamma+\ln 4\Bigr)
            \ . \label{eq:phisqr}
\end{eqnarray}
which is the same as Eq.~(\ref{eq:CR}).
Thus the result for $\langle 0|\phi^2|0\rangle$
is independent of the regularization schemes.

In relation to the usual definition of composite operators, the
subtractions in Eqs.~(\ref{eq:CR}) and (\ref{eq:phisqr})
correspond to additive renormalization\cite{BROWN_COL,BROWN89}
or mixing with the unit operator. The above simple exercise
suggests that the mixing of composite operators with the
poles in DR, or power terms $(\sim \mu^4,\ \mu^2)$ in CR,
are artifacts of defining
the singular part $S(x)$ to include these same terms.
To further assure this point we note the following.
The use of Eq.~(\ref{Dint}) with DR in
evaluating Feynman diagrams
can be justified in some cases by a careful consideration
of jacobian factors or extra
counterterms  proportional to a $\delta$-function
at the origin\cite{APFELDORF94}. In the OPE with DR, the validity
of  Eq.~(\ref{Dint}) is further related to the fact that
we apply the {\em same} regularization
and neglect similar integrals in the Wilson coefficients.
In the example of Eq.~(\ref{eq:pgtDR}), the use of DR to
regularize the infrared singularities in the first sum justifies
the use of Eq.~(\ref{Dint}) in calculating the integrals in
the second sum.
In other words, by using DR, we have implicitly and simultaneously
neglected power divergent terms in the Wilson coefficients and the
condensates. Thus neglecting the power terms in CR amounts to doing
the same thing implicit in DR. Note that the power terms
in the coefficients come from the infrared side of the momentum region,
while they appear in the condensate from the ultraviolet side.

In an asymptotically free theory
such as QCD, the genuine infrared
nonperturbative contributions,  in contrast to the power terms,
appear only in the condensates
if the coefficient functions are calculated in perturbation theory
as in the QCD sum-rule approach.  Thus the nonvanishing
gluon condensate is  physically significant and
not just a convenient definition, in contrast to
Ref.~\cite{NACHTMANN86}.
In the practical use of the OPE in QCD sum rules,
Wilson coefficients are calculated without the power dependence
\cite{SHIFMAN92}.
As a result the phenomenological condensates
as determined from the sum rules should not contain the power terms.
In fact, as will be shown below, if the power
terms were not excluded, they would be comparable to the
phenomenological gluon condensate at $\mu\sim 1\,$GeV, in
contradiction
with the sum-rule assumption of the dominance of
the genuine nonperturbative effects. Thus caution should be exercised
when one uses such phenomenological condensates
or makes comparison with them.

One might suspect that the above free-field exercise is too simple to
generalize, especially when there are infrared renormalon ambiguities.
In fact, there seems to be disagreement in the literature
\cite{NOVIKOV85,DAVID86}  on whether one can separate
the power terms from the genuine nonperturbative effects in the
two-dimensional O($N$) $\sigma$ model. Here we hope to clarify the situation.

We use the notation of Ref.~\cite{NOVIKOV85} and consider
$\langle \alpha^2\rangle^{(1)}\equiv\langle 0| \alpha^2|0\rangle-N m^4
$,
where $\alpha(x)$ is the auxiliary Lagrange-multiplier field
with vacuum expectation value $\sqrt{N} m^2$.
$\langle \alpha^2\rangle^{(1)}$
is the next-to-leading-order correction in $1/N$
and can be written as\cite{NOVIKOV85}
\begin{eqnarray}
\langle \alpha^2\rangle^{(1)}  & = & -4\pi
   \int_{k<\mu}\! {{\rm d}^2 k \over (2\pi)^2}\,
        \sqrt{k^2( k^2+4m^2)}\Big/\ln{\sqrt{ k^2+4m^2}+\sqrt{k^2}
              \over \sqrt{ k^2+4m^2}-\sqrt{k^2}}
                  \nonumber \\
   &=& -m^4\int_{1}^{A(\mu)}\! {\rm d}x\, {(x-1)^2\over x^2\ln x}
                 \ , \label{eq:sml}
\end{eqnarray}
where
\begin{equation}
A(\mu)=\biggl( \sqrt{1+{\mu^2\over 4m^2}}+{\mu \over 2m} \biggr)^4
             \ .
\end{equation}
We would like to point to the fact that the integrand in
Eq.~(\ref{eq:sml}) has no infrared singularities  even though its
large $k$ behavior contains a factor of the running coupling
 $f(k^2)=(4\pi)/\ln(k^2/m^2)$, which has an infrared pole
at $k^2=m^2$. As discussed
below, the QCD gluon condensate has a similar feature.
Not surprisingly, $\langle \alpha^2\rangle^{(1)}$ shows a power
divergence for large $\mu$.
Thus the integral in Eq.~(\ref{eq:sml}) has no way of producing any
ambiguous imaginary part in contrast to Ref.~\cite{DAVID86}.
A direct integration shows $\langle \alpha^2\rangle^{(1)}$
contains the exponential-integral function\cite{ABRAMOWITZ}
${\rm Ei}(\ln(\mu^4/m^4))$
for $\mu \gg m$. Here one does not make any assumption or prescription
in performing the integral and ${\rm Ei}(x)$ is a well-defined
principle-value integral with a known asymptotic behavior
\cite{ABRAMOWITZ}
\begin{equation}
{\rm Ei}(x)=-{1\over 2}\Bigl[{\rm E}_1(-x+i\epsilon)+
               {\rm E}_1(-x-i\epsilon)\Bigr]
          \sim e^{x}\sum_{n=0}^{\infty}{n!\over x^{n+1}}\ \ \ \
             (x\gg 1)
              \ . \label{eq:Ei}
\end{equation}
The only ambiguity that comes in is when all we know is the asymptotic
series in Eq.~(\ref{eq:Ei}) and we have to use a Borel transform to
resum it, in which case the  Borel transform has a singularity on the
positive real axis; this is the infrared renormalon. Note also that,
although the series in Eq.~(\ref{eq:Ei}) is not unambiguously
Borel summable due to
the infrared renormalon, it is still a useful asymptotic expansion
since its first $[x]$ terms, with $[x]$ meaning the integer part of
$x$,  is still a good approximation to ${\rm Ei}(x)$, as can be checked
numerically.
Thus if we allow power terms to include powers of $\mu$ multiplied
by an asymptotic series (Borel-summable or non-Borel-summable) in
the coupling constant, the separation of power terms from the
nonperturbative contributions is possible. In the present case,
the expansion of
$m^4 {\rm Ei}(\ln(\mu^4/m^4))$ provides an example of such  power terms.

Now we apply these ideas to the case of
the photon condensate
  \[\langle (\alpha / \pi) F^2\rangle \equiv
\langle 0 | (\alpha(\mu^2)/ \pi)F_{\mu\nu}F^{\mu\nu} |0\rangle \ ,\]
where $\alpha(\mu^2) = e^2/4\pi$ is the fine structure constant
renormalized at $\mu$ and $F_{\mu\nu}$ is
the electromagnetic field strength tensor. We note
an earlier discussion on the
photon condensate in Ref.~\cite{VAINSHTEIN89}.
We can define the photon condensate similar to Eq.~(\ref{eq:limit})
with $\phi(x)$ replaced by $F_{\mu\nu}(x)$. Note that this is a
gauge-invariant definition. With this definition, one
can  convince oneself that
\begin{equation}
\langle {\alpha \over \pi} F^2 \rangle =
        {2 \over \pi}\, \alpha (\mu^2) i\mu^{4-d}\!
       \int\!      {{\rm d}^d k \over (2\pi)^d}\,
     (k^2g^{\mu\nu}-k^{\mu}k^{\nu})D_{\mu\nu}(k)
                          \ , \label{eq:phcond}
\end{equation}
where some scheme-dependent subtraction is understood,
the momentum is Minkowskian with
the metric $g_{\mu\nu}=\text{diag}\,\text{(1 $-1$ $-1$ $-1$)}$,
and $D_{\mu\nu}(k)$
is the full photon propagator renormalized at $\mu$.
In a general covariant gauge
as represented by the parameter $\lambda$, current conservation
allows one to write~\cite{PASCUAL84}
\begin{equation}
D_{\mu\nu}(k)=\biggl [
  \biggl (-g_{\mu\nu}+{k_{\mu}k_{\nu} \over k^2-i\epsilon} \biggr )
   {1\over 1+\Pi(-k^2)}- {1\over \lambda}\,
     {k_{\mu}k_{\nu} \over k^2-i\epsilon} \biggr ]
  {1 \over k^2-i\epsilon}
                          \ , \label{eq:phpgt}
\end{equation}
where $\Pi(-k^2)$ is the photon polarization. Plugging
Eq.~(\ref{eq:phpgt}) into Eq.~(\ref{eq:phcond}) and Wick-rotating
to Euclidean space, we obtain
a  result independent of the gauge parameter,
\begin{equation}
\langle {\alpha \over \pi} F^2 \rangle =
        {6 \over \pi}\, \mu^{4-d}\!
       \int\!
      {{\rm d}^{d} k \over (2\pi)^{d}}\,
     {\alpha (\mu^2) \over 1+\Pi(k^2)}
                          \ . \label{eq:phcond1}
\end{equation}

Note that we cannot set the dimension of any integrals implicit in
the expression for $\Pi(k^2)$ to four before we perform the
explicit integration in Eq.~(\ref{eq:phcond1}).
Keeping $\Pi(k^2)$ at one-loop order and evaluating
Eq.~(\ref{eq:phcond1}) order-by-order in $\alpha(\mu^2)$
in DR, one finds
\begin{equation}
\langle {\alpha \over \pi} F^2 \rangle = 0 \ .
\end{equation}
However, it is instructive to perform the calculation by introducing
an ultraviolet Euclidean cutoff (or subtraction point) $\mu$.
To maintain gauge invariance,
we should use the result for $\Pi(k^2)$ from a gauge-invariant
renormalization. Also
power terms must be subtracted from the final result
as discussed earlier to compare with the result in DR.

Once we set $d=4$, the integrand in Eq.~(\ref{eq:phcond1})
becomes  the
running coupling constant $\alpha (k^2)$. Also, the integral
is dominated by the region of momentum
$k\sim \mu $, since the phase space provides a factor of $k^2$,
which suppresses the momentum region of very low momentum.
For $\mu \gg $ the electron mass,
we can then use the asymptotic form  of the
running coupling. For our purpose here, it suffices to keep only
the one-loop part which corresponds to summing up the
one-loop bubble diagrams. This is the well-known expression
\begin{equation}
     \alpha (k^2) =
    {\alpha(\mu^2) \over 1-[\alpha(\mu^2)/3\pi]
          \ln (k^2/{\mu}^2) }
        =  - {3\pi \over   \ln (k^2
                /\Lambda _{ \scriptscriptstyle\text{QED}}^2)}
                          \ , \label{eq:qedalp}
\end{equation}
where
$\Lambda _{ \scriptscriptstyle\text{QED}}^2
         ={\mu }^2e^{3\pi/\alpha (\mu^2)}$
is the location of the Landau ghost pole. After a Wick rotation,
the integral in Eq.~(\ref{eq:phcond1}) can be written in terms of the
exponential-integral function. For
$\mu \ll \Lambda _{ \scriptscriptstyle\text{QED}}$, it equals
\cite{ABRAMOWITZ}
\begin{equation}
      {9 \over 8\pi^2}\Lambda _{ \scriptscriptstyle\text{QED}}^4
     {\rm E}_1(\ln(\Lambda _{ \scriptscriptstyle\text{QED}}^4/\mu^4))
       ={3\mu^4 \over 16\pi^3}\sum_{n=0}^{\infty}
         {(-1)^n n!\over (6\pi)^n}[\alpha(\mu^2)]^{n+1}\ ,
\end{equation}
which are all power contributions and must be
subtracted to compare with the result of DR.
Thus we arrive again at a vanishing photon condensate.

We can deal similarly with
the gluon condensate of the physical QCD vacuum $|0\rangle$:
\begin{equation}
\langle {\alpha_{\rm s} \over \pi} G^2\rangle \equiv \langle 0 |
 {\alpha_{\rm s}(\mu^2) \over \pi}G_{\mu\nu}^aG^{\mu\nu a} |0\rangle
  \ , \ \ \ \ \ \ a=1, 2, \cdots, N_{\rm c}^2-1
\end{equation}
where $\alpha_{\rm s}(-k^2)=g^2(-k^2)/(4\pi)$ is the running strong
coupling and the  gluon field strength tensor is
\begin{equation}
     G_{\mu\nu}^a=
\partial _{\mu}A_{\nu}^a-\partial _{\nu}A_{\mu}^a
 +gf^{abc}A_{\mu}^bA_{\nu}^c \ ,
\end{equation}
and
\begin{equation}
     G_{\mu\nu}^aG^{\mu\nu a}=
       2(\partial _{\mu}A_{\nu}^a-\partial _{\nu}A_{\mu}^a)\,
        \partial ^{\mu}\!A^{\nu a}
      +4gf^{abc}\partial ^{\mu}\!A^{\nu a}A_{\mu}^bA_{\nu}^c
      +g^2f^{abc}f^{ade}A_{\mu}^bA_{\nu}^cA^{\mu d}A^{\nu e}
                         \ , \label{eq:Gsqr}
\end{equation}
with $f^{abc}$ the structure constants of
the $SU(N_{\rm c})$ color group.
In contrast to the photon condensate, a definition
analogous to
Eq.~(\ref{eq:limit}) may not apply since
$T[G_{\mu\nu}^a(x)G^{\mu\nu a}(0)]$ is not
gauge invariant.
Furthermore, mixing between the $G^2$ operator and gauge-variant
operators must be taken into account if a gauge with ghosts is chosen
\cite{GROSS74}.
Nevertheless, it is tempting to generalize
the QED result in Eq.~(\ref{eq:phcond1}) to obtain%
\footnote{We note some estimates of the gluon condensate from the
gluon propagator in Ref.~\cite{ROBERTS92}.}
\begin{equation}
\langle {\alpha_{\rm s} \over \pi} G^2\rangle
     =       {6 \over \pi}\, (N_{\rm c}^2-1)\,\mu^{4-d}\!
       \int {{\rm d}^d k \over (2\pi)^d}\,
         \alpha _{\rm s}(k^2)
           \ , \label{eq:qcdcond}
\end{equation}
where the $(N_{\rm c}^2-1)$ factor comes from the sum over colors.
Note that Eq.~(\ref{eq:qcdcond}) involves the running coupling
at any given scale to all orders and the leading-order term coincides
with the result in Ref.~\cite{NOVIKOV85}.

To give some justification for
Eq.~(\ref{eq:qcdcond}), we can include the $G^2$ operator
in the QCD lagrangian with a constant Lagrange
multiplier $\xi$ and choose a physical gauge such as the
axial gauge with gauge fixing term
\begin{equation}
   -{1\over 2\lambda}(n\cdot A^a)^2 \ ,
\end{equation}
where $n_{\mu}$ is some arbitrary fixed vector. Since the ghosts
decouple\cite{LEIBBRANDT87,ACERBI94}, there is no operator mixing with
the gluon operator in the chiral limit.
After evaluating the usual connected generating
functional as a function of $\xi$, we can obtain the gluon
condensate by taking a derivative at  $\xi=0$. Note that it is
convenient if one rescales the fields by letting
$A\rightarrow A/\sqrt {1+\xi}$, and similarly for $g$ and $\lambda$.
This way one can show that,
at least at one-loop order, the gluon condensate
is independent of the gauge parameter because
$g^2\lambda$ is not affected by the above scaling. The result turns out
to be  the same as that obtained from using Eq.~(\ref{eq:limit}) with
$\phi(x)$ replaced with $G_{\mu\nu}^a(x)$ and
gauge-parameter-dependent terms disregarded. As a result,
the first term of Eq.~(\ref{eq:Gsqr}) alone will produce the result
of Eq.~(\ref{eq:qcdcond}) at least at one-loop order
if one uses the known one-loop gluon polarization
in physical gauges\cite{LEIBBRANDT87,DALBOSCO86}.
Note that, in perturbation theory,
one can show  order-by-order
in dimensional regularization that all the terms in
Eq.~(\ref{eq:Gsqr}) yield vanishing contributions to the condensate,
although the
nonperturbative sum to all orders may not vanish. Intuitively, we
expect that the first term of Eq.~(\ref{eq:Gsqr}) should
dominate the nonperturbative  gluon condensate since it
is a one-body operator and its matrix element involves only
the creation and absorption of a single gluon while the matrix
elements of the rest of the
terms involve two or more gluons.
Thus we conjecture that Eq~(\ref{eq:qcdcond}) is
a good approximation in QCD.

It is interesting to see what we can learn from
Eq.~(\ref{eq:qcdcond}) for the perturbative QCD vacuum, which is here
defined  by requiring the
running coupling to be given by the asymptotic-freedom result.
At the one-loop level, we have
\begin{equation}
\alpha_{\rm s}(k^2) = {4\pi \over \beta_{\scriptscriptstyle 0}
   \ln ( (k^2-i\epsilon)/\Lambda_{\scriptscriptstyle\text{QCD}}^2 ) }
               \ , \label{eq:qcdalp}
\end{equation}
where $\beta_{\scriptscriptstyle 0}=(11N_{\rm c}-2N_{\rm f})/3$
is the first coefficient of the $\beta$-function of QCD with $N_{\rm f}$
light quarks and $\Lambda_{\scriptscriptstyle\text{QCD}}$ is the
QCD scale parameter. Note that we have retained the Feynman $i\epsilon$
prescription even after the Wick rotation to Euclidean space.

We again use a cutoff regularization with
$\mu > \Lambda_{\scriptscriptstyle\text{QCD}}$.
In earlier work, integrals similar to Eqs.~(\ref{eq:qcdcond})
and (\ref{eq:qcdalp})
have been examined to show the factorial growth of the
coefficients of perturbation theory\cite{BIGI94}.
Here we note that our specification of
the Feynman $i\epsilon$ prescribes the way
to go around the infrared singularity and allows to perform the
integral in Eq.~(\ref{eq:qcdcond}) analytically.
The principal value of the integral gives a real part that can again
be written as an exponential-integral function and integration
around the infrared singularity leads to an imaginary part.
We find
\begin{equation}
\langle {\alpha_{\rm s} \over \pi} G^2\rangle_{\rm pert} =
     i {3(N_{\rm c}^2-1) \over 2\pi \beta_{\scriptscriptstyle 0}}\,
         \Lambda_{\scriptscriptstyle\text{QCD}}^4
                 \ , \label{eq:result}
\end{equation}
where we have subtracted the power terms:
\begin{equation}
      {3(N_{\rm c}^2-1) \over 2\pi^2 \beta_{\scriptscriptstyle 0}}\,
         \Lambda_{\scriptscriptstyle\text{QCD}}^4
    {\rm Ei} (\ln (\mu^4/ \Lambda_{\scriptscriptstyle\text{QCD}}^4))
    ={3\mu^4 \over 16\pi^3}(N_{\rm c}^2-1)\sum_{n=0}^{\infty}
              n!\Bigl({\beta_0\over 8\pi} \Bigr)^n
             [\alpha_{\rm s}(\mu^2)]^{n+1} \ . \label{eq:QCDpowers}
\end{equation}
Thus the energy
density of the perturbative vacuum is imaginary, indicating its
instability.

One may ask: where are the ambiguities of the infrared
renormalons? As discussed earlier,
the renormalon ambiguities appear only when one tries to resum
the asymptotic perturbation series
in Eq.~(\ref{eq:QCDpowers}) using a Borel
transform. The Feynman $i\epsilon$ prescription gives the same
result as that obtained from the Borel summation method when
the integration contour in the Borel plane is deformed to pass
above the infrared renormalon pole.
Thus the Feynman $i\epsilon$ prescription
says unambiguously that the perturbative QCD vacuum is unstable
as expected.
Notice also that, had we considered the power terms in
Eq.~(\ref{eq:QCDpowers}) as the perturbative contribution to the
phenomenological gluon condensate,
at $\mu \sim 1\,$GeV this contribution would dominate
the phenomenological value of $\sim (350\, {\rm MeV})^4$
determined from QCD sum rules.
In Ref.\cite{NOVIKOV85}, the estimated power contribution is
small because a low scale $\mu$ is used.

At present we know little about the physical
running coupling constant of QCD in the low-momentum region, although
we expect  $\alpha _{\rm s}(k^2)$ to exhibit
asymptotic freedom for large momentum. However,
given the phenomenological value of the gluon
condensate from QCD sum rules, we can consider
Eq.~(\ref{eq:qcdcond}) as a constraint for the running coupling.
Such a constraint may be important in deciding what Ansatz to use
for the running coupling in some QCD phenomenology such as the
DSE approach.

In the literature, the  Ans\"{a}tze used in the
DSE approach are a combination of the following:
\begin{eqnarray}
   \alpha_0(k^2) &=& {4\pi \over \beta_{\scriptscriptstyle 0}
   \ln (\tau +k^2 /\Lambda_{\scriptscriptstyle\text{QCD}}^2 ) }
        \ , \\[4pt]
   \alpha_1(k^2) &=& \pi ak^4e^{-k^2/\Delta}
               \ , \\[4pt]
   \alpha_2(k^2) &=& {3\pi \chi^2\over 4\Delta^2} k^2e^{-k^2/\Delta} \  ,
\end{eqnarray}
where $\tau$, $a$, $\Delta$, and $\chi$ are phenomenological
parameters. These yield the following corresponding contributions to the
condensate:
\begin{eqnarray}
\langle {\alpha_{\rm s} \over \pi} G^2\rangle_0 &=&
      {3(N_{\rm c}^2-1) \over 2\pi^2 \beta_{\scriptscriptstyle 0}}\,
         \Lambda_{\scriptscriptstyle\text{QCD}}^4
  \left[        \tau \,   {\rm Ei} (\ln \tau)
           - {\rm Ei}  \left(\ln \tau^2\right) \right ] \ ,
                     \label{eq:G0}\\[4pt]
\langle {\alpha_{\rm s} \over \pi} G^2\rangle_1 &=&
     {9(N_{\rm c}^2-1) \over 4\pi^2 }\,
       a\Delta^4                     \ ,\label{eq:G2} \\[4pt]
\langle {\alpha_{\rm s} \over \pi} G^2\rangle_2 &=&
     {9(N_{\rm c}^2-1) \over 16\pi^2 }\,
       \chi^2\Delta                   \ . \label{eq:G3}
\end{eqnarray}
Notice that
$ \alpha_0(k^2)$ is an extension of the one-loop
renormalization group result to low-momentum region;
typically $\tau$ is taken to be $2\ {\rm to}\ 10$.
To obtain Eq.~(\ref{eq:G0}),
we have subtracted power terms, which are the same as
Eq.~(\ref{eq:QCDpowers}) for
$\mu^2 \gg \Lambda_{\scriptscriptstyle\text{QCD}}^2 $.
For $N_{\rm c}=N_{\rm f}=3$ and $\tau=3$, we find
$\langle (\alpha_{\rm s}/ \pi) G^2\rangle_0
     =(0.568 \Lambda_{\scriptscriptstyle\text{QCD}})^4$, while,
for $\tau=10$,
$\langle (\alpha_{\rm s}/ \pi) G^2\rangle_0
     =(1.44 \Lambda_{\scriptscriptstyle\text{QCD}})^4$.
In Ref.~\cite{JAIN93}, the running coupling is taken to be
$\alpha_{\rm s}(k^2)=\alpha_0(k^2)+\alpha_1(k^2)$  with
$a=(387\, {\rm MeV})^{-4}$, $\Delta=(510\, {\rm MeV})^2$,  $\tau=10$,
and $\Lambda_{\scriptscriptstyle\text{QCD}}=228\,$MeV. The resulting
gluon condensate is  $(788\, {\rm MeV})^4$.
In Ref.~\cite{PRASCHIFKA89}, the running coupling is
$\alpha_{\rm s}(k^2)=\alpha_0(k^2)+\alpha_2(k^2)$  with
$\chi= 1.14\,$GeV, $\Delta=0.002\ {\rm GeV}^2$, $\tau=3$,
and $\Lambda_{\scriptscriptstyle\text{QCD}}=190\,$MeV. This gives
a gluon condensate of $(191\, {\rm MeV})^4$.
Thus both results differ significantly from the value determined
from QCD sum rules. Nevertheless, the parameters are somewhat
flexible, as indicated by the authors in Ref.~\cite{ROBERTS92}.
Thus our result provides a further
constraint on the parameters.

Finally, we remark that
the gluon condensate as given by
Eq.~(\ref{eq:qcdcond}) is not sensitive to the behavior of
$\alpha _{\rm s}(k^2)$ near the origin because of
phase-space suppression. Thus, if confinement is associated with
the running coupling near the origin, our result indicates that
the nonzero gluon condensate is not neccessarily
related to confinement (see also Ref.~\cite{ROBERTS92}).

To reiterate, we have argued that the power
terms can be
excluded consistently from the OPE. We have shown that the
photon condensate vanishes and suggested that the gluon condensate can be
simply related to the physical running coupling of QCD. Finally,
we examine some Ans\"{a}tze for the running coupling that have
appeared in the literature. The resulting values for the
gluon condensate are  different from
the value  determined from QCD sum rules.

\acknowledgments

We are pleased to thank X. Ji, C. D. Roberts, D. G. Robertson,
and R. J. Perry  for useful conversions and comments.
This work was supported in part by the National Science
Foundation
under Grant Nos.~PHY-9203145, PHY-9258270, PHY-9207889, and
PHY-9102922, and the Sloan Foundation.

\end{document}